\documentclass[twocolumn,
showpacs,preprintnumbers,
showkeys,
amsmath,amssymb,
prl,reprint,
superscriptaddress
]{revtex4-1}

\usepackage{graphicx}
\usepackage{dcolumn}
\usepackage{bm}
\usepackage{hyperref}
\usepackage{textcomp}
\usepackage{changes}

\begin{document}


\title{Carrier drift-control of spin currents in graphene-based spin current demultiplexers}

\author{J. Ingla-Ayn\'es}
 \email{josepingla@gmail.com}
 \affiliation{Physics of Nanodevices, Zernike Institute for Advanced Materials, University of Groningen, The Netherlands}

\author{A. A. Kaverzin}
\affiliation{Physics of Nanodevices, Zernike Institute for Advanced Materials, University of Groningen, The Netherlands}
\author{B. J. van Wees}

\affiliation{Physics of Nanodevices, Zernike Institute for Advanced Materials, University of Groningen, The Netherlands}

\date{\today}

\begin{abstract}
Electrical control of spin transport is promising for achieving new device functionalities. Here we calculate the propagation of spin currents in a graphene-based spin current demultiplexer under the effect of drift currents. We show that, using spin and charge transport parameters already achieved in experiments, the spin currents can be guided in a controlled way. In particular, spin current selectivities up to $10^2$ can be achieved when measuring over a 10~\textmu m distance under a moderate drift current density of 20~\textmu A/\textmu m, meaning that the spin current in the arm which is `off' is only 1\% of the current in the `on' arm. To illustrate the versatility of this approach, we show similar efficiencies in a device with 4 outputs and the possibility of multiplexer operation using spin drift. Finally, we explain how the effect can be optimised in graphene and 2D semiconductors.  
 
\end{abstract}

\maketitle

\section{Introduction}

The ability to manipulate spin currents by electrical means is a major requirement to achieve functional spintronic devices \cite{Fabian_Acta}. For this purpose, graphene is an ideal candidate as a transport media thanks to its superior spin and charge transport properties\cite{RevModPhysGraph,NikoSpinGraf,ReviewFabian,FlagshipReview,12nsAachen}. These properties have allowed for room temperature spin relaxation lengths up to 30~\textmu m in an experiment where spin transport is diffusive \cite{12nsAachen} and up to 90~\textmu m when carrier drift was additionally induced in the channel \cite{DriftNanoLetters}. Moreover, enhancements in the spin injection and detection efficiencies \cite{Gurram2017, SrO2Kawakami, Neumann2013} show the possibility to create unprecedentedly large spin accumulations in graphene, resulting in larger signals useful for future spintronic operations.\\
 It has been shown that one can perform logic operations using spins in graphene in the diffusive regime. In particular, the interplay between the spin injection efficiencies of different contacts can be used to perform logic XOR operations \cite{XORKawakami}. The introduction of transition metal dichalcogenides (TMDs) also allows for control of the spin signals in graphene-based spintronic devices. The tunability of the TMD resistance allows for efficient control of the spin absorption in such structures resulting in `on/off' ratios up to 60 \cite{yan2016,dankert2017}. Another means of controlling spin currents was demonstrated in Y-shaped graphene devices via the manipulation of carrier densities in the different arms individually \cite{MuxFert}. This approach leads to spin guiding thanks to the change in spin lifetime and resistivity of graphene with the carrier density. On-off ratios for spin currents up to 7 were obtained when assuming that the spin relaxation time in the channel decreases when increasing the resistivity, a condition only achieved experimentally for bilayer graphene on SiO$\mathrm{_2}$ at temperatures equal or lower than 50~K \cite{BLGAachen,BLGKawakami,avsar2011}.\\
 The spin signals achieved using diffusive transport are limited by the long diffusion times, that increase with the square of the distance \cite{Fabian_Acta}. However, a charge current can significantly influence the spin propagation. The average velocity of spin carriers becomes much larger (smaller) when transport occurs in the direction of (against) the drift velocity ($\mathrm{v_d}$). This allows for efficient control of the spin relaxation length using charge currents \cite{DriftNanoLetters, DriftGraphenePRL} without affecting the spin lifetime in the system. In addition, $\mathrm{v_d}$ is inversely proportional to the carrier density. This brings a new mechanism to control the magnitude and sign of $\mathrm{v_d}$. 
 The tunability of spin transport with drift enables for new device functionalities, such as demultiplexers, that are devices that route the input signal to an output that is controlled by the select line \cite{HandbookElec}. Demultiplexers have applications ranging from analog switches in high frequency transmission lines to decoders in digital electronics \cite{Network,HandbookIBM} and combination of spin currents with multiplexer operations could allow for `in situ' memory capabilities enabling for new functionalities \cite{behin2010,behin2011}.\\  
 Here we show that, by applying drift currents in a Y-shaped geometry, the spin currents can be controlled in an extremely efficient way, realizing the demultiplexer functionality. Our calculations using the drift-diffusion equations show that one can achieve spin current selectivities $\mathrm{I_{top}/I_{bot}}$ (where $\mathrm{I_{top(bot)}}$ is the spin current in the top (bottom) arm in Fig. \ref{fig:figure-1}) as high as $10^3$ for drift current densities of 100~\textmu A/\textmu m when measuring close to the bifurcation. At a distance $\mathrm{L}=$~10~\textmu m from the bifurcation, the selectivities increase up to $10^6$, a value that stays as high as $10^2$ for drift current densities of 20~\textmu A/\textmu m.  Moreover, we use the same model for a geometry constituting of 2 Y-shaped graphene channels connected to the output of another Y-shaped graphene channel, and obtain similar performances of $\mathrm{10^2}$ for drift currents of 25~\textmu A/\textmu m. We also explain how the effect of drift can be used to achieve multiplexer operation. From these considerations we conclude that the introduction of drift leads to several advantages when compared with the diffusive case. These advantages include a significant enhancement in the speed of operations, a significant improvement of the `on/off' ratio, and higher spin currents due to the increase of the spin relaxation lengths induced by the drift. On the other hand, the introduction of drift to spin logic also causes extra power consumption induced by the drift currents. We argue that this can be greatly reduced by replacing graphene with a semiconducting 2D material.
 
 Moreover, the technique described here can also be used in combination with other implementations, such as the XOR operation described in \cite{XORKawakami}, that can be added to our geometry to combine multiplexer with logic operations.

\section{The model}
To determine the spin currents and spin accumulations in the different geometries studied here under the effect of drift, we use the drift-diffusion equations \cite{YuFlatte}.\\
Since in most graphene-based spin valve devices the contacts inject spins homogenously over the channel width, we reduce the spin propagation to a one-dimensional problem. Also, we assume that the channel width ($\mathrm{W_s}$) is much longer than the mean free path. This condition is commonly achieved for all graphene spintronic devices studied at room temperature for $\mathrm{W_s=1}$~\textmu m (the mean free path is 0.16~\textmu m in our case). This condition makes our results independent of the exact geometry of the bifurcation. We also assume that the contacts are not invasive, this condition is achieved when the contact resistances are much higher than the channel resistance \cite{ElectronSpinPopinciuc}.\\
When a charge current $\mathrm{I_d}$ is applied to a non-magnetic channel to induce drift, the spin current, which is defined as $\mathrm{I_s=I_\uparrow-I_\downarrow}$, where $\mathrm{I_{\uparrow(\downarrow)}}$ is the current of up (down) spins, propagating in the channel is \cite{IvanRapCom}:
\begin{equation}
\mathrm{
I_s=\frac{W_s}{eR_{sq}}\left(-\frac{d\mu_s}{dx}+\frac{v_d}{D_s}\mu_s
\right)
}
\label{JsDef}
\end{equation}
Here $\mathrm{D_s}$ is the spin diffusion coefficient, $\mathrm{R_{sq}}$ is the square resistance, $\mu_s$ is the spin accumulation, and $\mathrm{e}$ is the electron charge. The drift velocity is defined as $\mathrm{v_d=\mu I_dR_{sq}/W_s=I_d/(enW_s)}$, where $\mathrm{n}$ is the carrier density and $\mu$ is the electron mobility. The first term in Eq.~\ref{JsDef} describes the spin diffusion and the second one describes the spin current induced by the pulling of the spin accumulation induced by the drift.\\
 $\mu_s$ in the channel follows the drift-diffusion equations \cite{YuFlatte}:
\begin{equation}
\mathrm{
D_s\frac{d^2\mu_s}{dx^2}+v_d\frac{d\mu_s}{dx}-\frac{\mu_s}{\tau_s}=0
}
\label{DriftDiffEq}
\end{equation} 
 Here $\tau_s$ is the spin lifetime. This equation has solutions of the form $\mathrm{\mu_s=A\exp(x/\lambda_+)+B\exp(-x/\lambda_-)}$ where the coefficients $\mathrm{A}$ and $\mathrm{B}$ are determined by both the device geometry and the spin relaxation lengths, which are:
\begin{equation}
\mathrm{
\lambda_\pm^{-1}=\pm \frac{v_d}{2D_s}+\sqrt{\left(\frac{v_d}{2D_s}\right)^2+\frac{1}{\lambda^2}}
}
\label{SpinRelaxationLengths}
\end{equation}
Here $\mathrm{\lambda=\sqrt{D_s\tau_s}}$ is the spin diffusion length in the channel.
 $\lambda_+$ and $\lambda_-$ are the so called upstream and downstream spin relaxation lengths. They describe spin transport opposed to and along with the drift velocity respectively. Their difference provides an asymmetry in the spin propagation, which is the source of spin current selectivity in our calculations.\\
We use the solutions of Eq. \ref{DriftDiffEq} to describe the spin accumulation in the different parts of the sample. Because the drift currents and/or carrier densities are different in the different parts of the sample, the unknown coefficients are obtained using the following boundary conditions taking into account the device geometry: $\mu_s$ is zero infinitely far away from the injector and it is continuous in the entire device, including the boundaries between the different regions. The spin current $\mathrm{I_s}$ is conserved at the different junctions apart from the injection point. The spin injector is a ferromagnetic electrode placed at the bifurcation point and it induces a discontinuity in the spin current of $\mathrm{P_i I_{inj}/e}$. $\mathrm{P_i}$ is the spin polarization of the injector and $\mathrm{I_{inj}}$ is the charge current applied to the injector electrode. For simplicity, the additional electrodes that are
used to induce drift are placed further away from the bifurcation points than the relevant spin relaxation length and, therefore, are not considered to be spin injectors irrespective of the constituting materials. Under these conditions, we are able to obtain the spin currents and nonlocal resistances $\mathrm{R_{s}}=\mathrm{V_{s}/I_{inj}=\mu_sP_d/(eI_{inj})}$ (where $\mathrm{V_s}$ $\mathrm{=\mu_sP_d/e}$ is the voltage drop caused by the spin accumulation at the detecting ferromagnetic electrode that has a spin polarization $\mathrm{P_d}$) in the reported device geometries \cite{ElectronSpinPopinciuc,MarcosSuspended}. \\
Because of the inverse dependence between $\mathrm{v_d}$ and $\mathrm{n}$, $\mathrm{v_d}$ theoretically diverges at the neutrality point. However, thermal broadening \cite{EHPuddles, DiracPointGraphene} limits the minimal electron and hole densities down to a value of $\mathrm{n_{th}^e=n_{th}^h=\pi/12(k_BT/(\hbar v_f))^2=4\times 10^{14}}$~$\mathrm{m^{-2}}$ at room temperature. In a system with equal amount of electrons and holes no drift is present since the drift velocities for electrons and holes are opposite. Thus, an imbalance between the electron and hole densities is required to achieve a non-zero average drift velocity. This condition increases the minimal carrier density that is optimal for spin drift and, therefore, limits the $\mathrm{v_d}$ that can be reached experimentally.
  For all the calculations reported here, we have used transport parameters which have been achieved experimentally in monolayer graphene \cite{12nsAachen} near the charge neutrality point. They can be found in table \ref{table1}. In our case, the optimal carrier density is of 8$\mathrm{\times 10^{15}~m^{-2}}$ and, for a drift current density of 100~\textmu A/\textmu m, gives a drift velocity $\mathrm{v_d=8\times 10^4}$~m/s.  This value is still 1 order of magnitude smaller than the maximum drift velocity achieved for boron nitride encapsulated graphene, which is around $0.55\times10^6$~m/s at room temperature for a carrier density of about 9$\times10^{15}$~$\mathrm{m^{-2}}$\cite{DriftVelSatEncGr} \footnote{Correcting for the saturation in $\mathrm{v_d}$ leads to a drift velocity of 7.4$\times10^4$~m/s, a correction of only 7.5\%. For the highest drift current reported here, $\mathrm{I_d=400}$~\textmu A/\textmu m $\mathrm{v_d=3.2\times 10^5}$~m/s, that after correction becomes $\mathrm{2.3\times 10^5}$~m/s, a correction of 28\%}.
 \begin{table}[t]
    \caption{Spin, charge transport parameters, contact spin polarizations estimated from \citep{12nsAachen} and channel width. For simplicity we assume that the system is electron-hole symmetric and the sign of the drift velocity changes for hole transport.}
        \begin{ruledtabular}
        \begin{tabular}{c c c c c c c}
            \renewcommand{\arraystretch}{2}
            $\mathrm{\mu}$($\mathrm{m^2/Vs}$) & $\mathrm{R_{sq}(\Omega)}$&$\mathrm{n(m^{-2})}$& $\mathrm{\tau_s(ns)}$ &
             $\mathrm{D_s(m^2/s)}$&$\mathrm{P_i=P_d}$& $\mathrm{W_s}$(\textmu m)\\
             1 & 800 &8$\times 10^{15}$& 4 &0.08&0.1&1 \\
             
    \end{tabular}
    \end{ruledtabular}
    \label{table1}
\end{table}

\section{Results}
\subsection{Geometry I}
\begin{figure}[tb]
	\centering
		\includegraphics[width=0.4\textwidth]{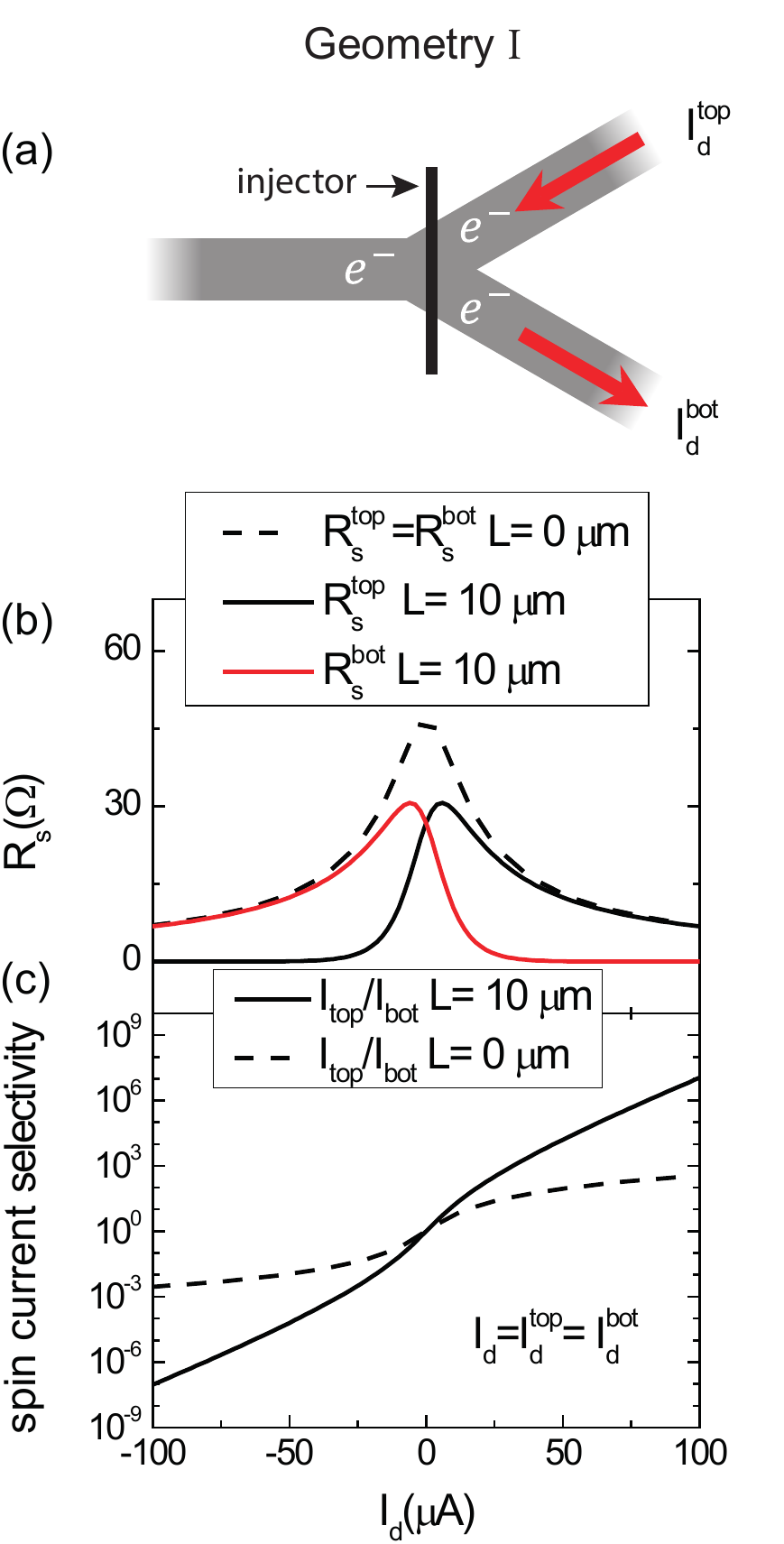}
	\caption{(a) Sketch of device geometry I. The red arrows represent the charge current direction for positive $\mathrm{I_d}$. The carriers are electrons in all 3 arms, that are assumed to be infinitely long. (b) $\mathrm{R_{\added{s}}}$ vs $\mathrm{I_d}$ in the top and bottom arms and close to the bifurcation and at a 10~\textmu m distance. (c) Spin current selectivity calculated both close to the bifurcation ($\mathrm{L=0}$) and at $\mathrm{L=10}$~\textmu m.}
	\label{fig:figure-1}
\end{figure}
We start the discussion of our results obtained for the Y-shaped device geometry, where the spin injector is placed at the bifurcation point. For geometry I, the 3 arms have the same transport parameters. The spins are guided by applying opposite drift currents in the top and bottom arms, as shown in Fig.~\ref{fig:figure-1}, and generate a spin signal $\mathrm{R_{s}^{top(bot)}}$ at the top (bottom) arm. In the left arm there is no net charge current and the propagation of spins is determined by diffusion. As we can see in Fig \ref{fig:figure-1}b, when a drift current is applied, $\mathrm{R_{s}}$ close to the bifurcation point decreases. This is because the spins are pulled away from the injector towards the top or bottom arm, reducing the spin accumulation. For the ratio $\mathrm{I_{top}/I_{bot}}$ at $\mathrm{L=0}$ we observe that, when the applied drift current is of 100~\textmu A, $\mathrm{I_{top}/I_{bot}}$ is as high as $10^3$. This implies that 99.9\% of the spins are propagating along with the drift velocity and only 0.1\% of them propagate to the opposite arm. Looking at the spin signal 10~\textmu m away from the bifurcation we see that the nonlocal resistance in the top arm differs from that in the bottom arm. This difference develops very rapidly with the applied drift due to both the exponential decay of the spin current with the distance and the difference between upstream and downstream spin relaxation lengths. For example, for $\mathrm{I_d=100}$~\textmu A, the spin current selectivities become as high as $10^6$, with $\lambda_-=340$~\textmu m and $\lambda_+=0.97$~\textmu m. We also note that both $\mathrm{R_s^{top}}$ and $\mathrm{R_s^{bot}}$ at L=10~\textmu m decrease for drift currents higher than $\pm$6~\textmu A. This is because, for infinitely long arms, the drift spreads the spins over a large distance $\lambda_-$. Since the injected spin current is constant, this results in reduction of the spin accumulation close to the injector. This effect can be reduced by applying the drift over a finite length of the channel \cite{IvanRapCom}.\\
 In real device applications, the power consumption has to be kept minimal and, to achieve this, we determine the device performance for lower drift currents. At $\mathrm{I_d=20}$~\textmu A a current selectivity of $10^2$ is achieved. We consider this value to be enough for basic operations.
It is also worth noting that, when applying a drift current of 100~\textmu A, 4.8\% of the spin current is still propagating to the left arm, a value that goes up to 20\% when $\mathrm{I_d=20}$~\textmu A.  
\subsection{Geometry II}
\begin{figure}[tb]
	\centering
		\includegraphics[width=0.4\textwidth]{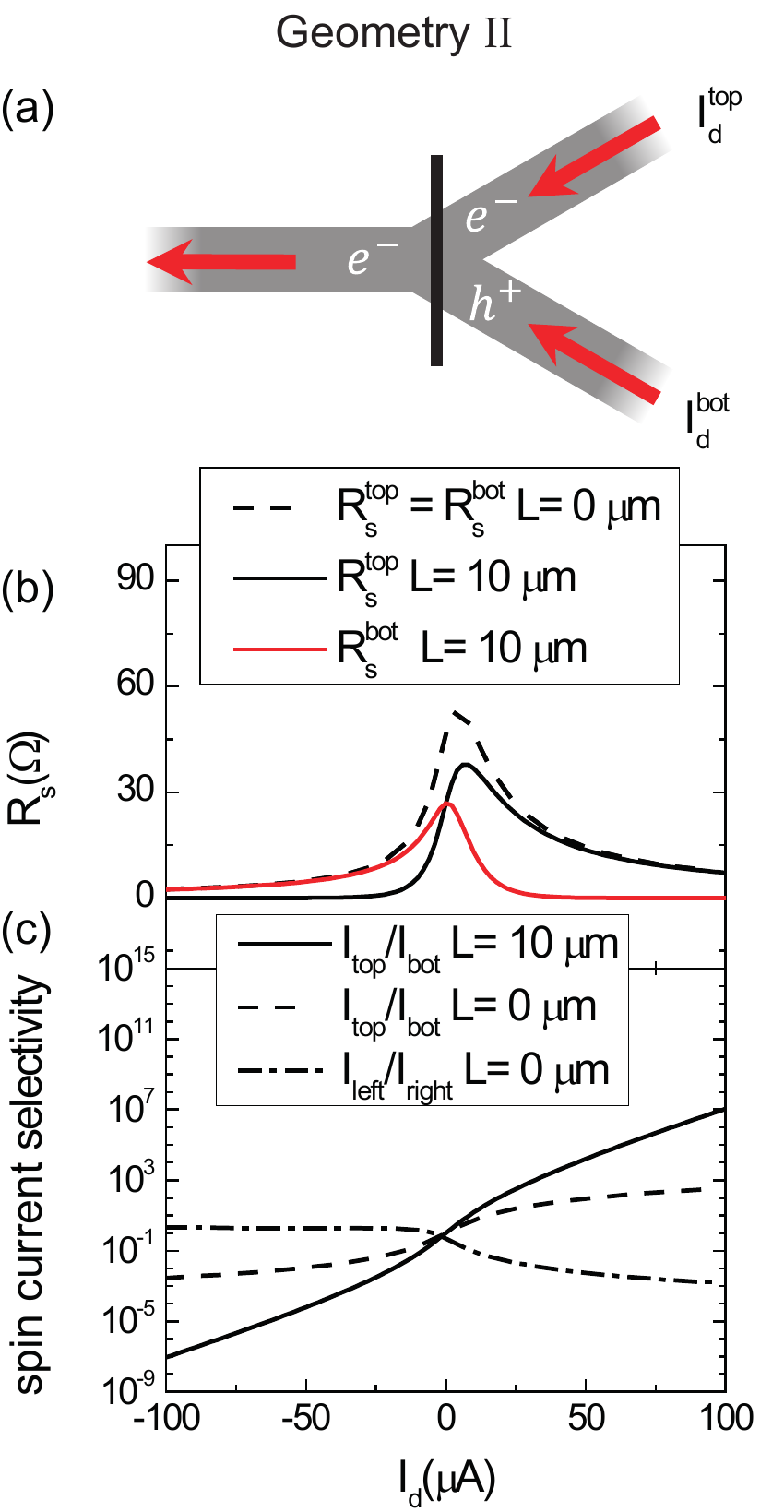}
	\caption{(a) Sketch of device geometry II, the red arrows represent the charge current direction for positive $\mathrm{I_d}$. Note that the drift velocity for electrons oposes the charge current direction. The carriers are holes in the bottom arm and electrons in the others. All the arms are assumed to be infinitely long. (b) $\mathrm{R_{s}}$ vs $\mathrm{I_d}$ in the top and bottom arms and at the bifurcation and at a 10~\textmu m distance. (c) Spin current selectivity calculated both at the bifurcation and at a 10~\textmu m distance.}
	\label{fig:figure-2}
\end{figure}

As mentioned in the previous section, $\mathrm{I}_\mathrm{left}/\mathrm{I}_\mathrm{right}$, which is defined as the spin current propagating to the left arm $\mathrm{I}_\mathrm{left}$ normalized by $\mathrm{I}_\mathrm{right}=\mathrm{I}_\mathrm{top}+\mathrm{I}_\mathrm{bot}$, is determined by diffusion in Geometry I and it can be reduced by applying a net drift current in the left arm. This is achieved by changing the carrier type of either the top or the bottom arm depending on the selected output.
When one changes the carriers from electrons to holes, the drift velocity  reverses. This allows us to apply the drift current in both top and bottom arms in the same direction ($\mathrm{I_d^{top}=I_d^{bot}=I_d}$) while the drift velocities are opposite ($\mathrm{v_d^{top}}=\mathrm{-v_d^{bot}}$). In this case, the drift current in the left arm is non zero and equal to $2\times \mathrm{I_d}$. As shown in  Fig. \ref{fig:figure-2}b, $\mathrm{R_{s}}$ at L=0 is no longer symmetric with respect to $\mathrm{I_d}$. This is caused by the spin drift in the left arm, that blocks spin propagation to the left at positive drift currents but promotes spin propagation in this arm when $\mathrm{I_d}$ is negative. 

As a consequence, we observe that the most efficient operation of this device (maximum $\mathrm{I_{top}/I_{bot}}$ and minimum $\mathrm{I}_\mathrm{left}/\mathrm{I}_\mathrm{right}$) occurs when the drift current is positive (drift currents in the direction of the red arrows in Fig.~\ref{fig:figure-2}a). The output terminal, where the current is directed, can be controlled by changing the carrier densities in the top and bottom arms while keeping the left arm at the same density and applying positive drift currents (in the direction of the red arrows, Fig.~\ref{fig:figure-2} a). The spin current selectivities comparing top and bottom arms are the same as for geometry I due to electron-hole symmetry. The difference is that the spin current propagating to the left arm is kept minimal. In particular, when $\mathrm{I_d=100}$~\textmu A, only 0.14\% of the injected spin currents propagate into this arm, a value that stays as low as 3\% for $\mathrm{I_d=20}$~\textmu A, confirming the feasibility of device operation with moderate drift currents.\\
\subsection{Demultiplexing operation}
As shown above, the most efficient way of controling the spin demultiplexing operation is via the carrier type (density). To describe the practical operation of device geometry II, we write a truth table (table \ref{table2}) by defining the inputs as the gate voltages that have to be applied to each separate arm to control the densities. We call $\mathrm{V_{gL}}$, $\mathrm{V_{gT}}$, $\mathrm{V_{gB}}$ the left, top, and bottom arm gates respectively and define them as `0' when the carriers in the channel are holes and `1' when they are electrons.
   \begin{table}[b]
    \caption{Truth table for the 2 leg demultiplexer operation of geometry II for positive $\mathrm{I_d}$ (Fig. \ref{fig:figure-2})}
        \begin{ruledtabular}
        \begin{tabular}{c c c c c }
            \renewcommand{\arraystretch}{2}
            $\mathrm{V_{gL}}$& $\mathrm{V_{gT}}$&$\mathrm{V_{gB}}$& $\mathrm{I_{top}}$& $\mathrm{I_{bot}}$\\
            \hline
1&1&0&1&0\\
1&0&1&0&1\\
             
    \end{tabular}
    \end{ruledtabular}
    \label{table2}
\end{table}

\begin{figure}[tb]
	\centering
		\includegraphics[width=0.4\textwidth]{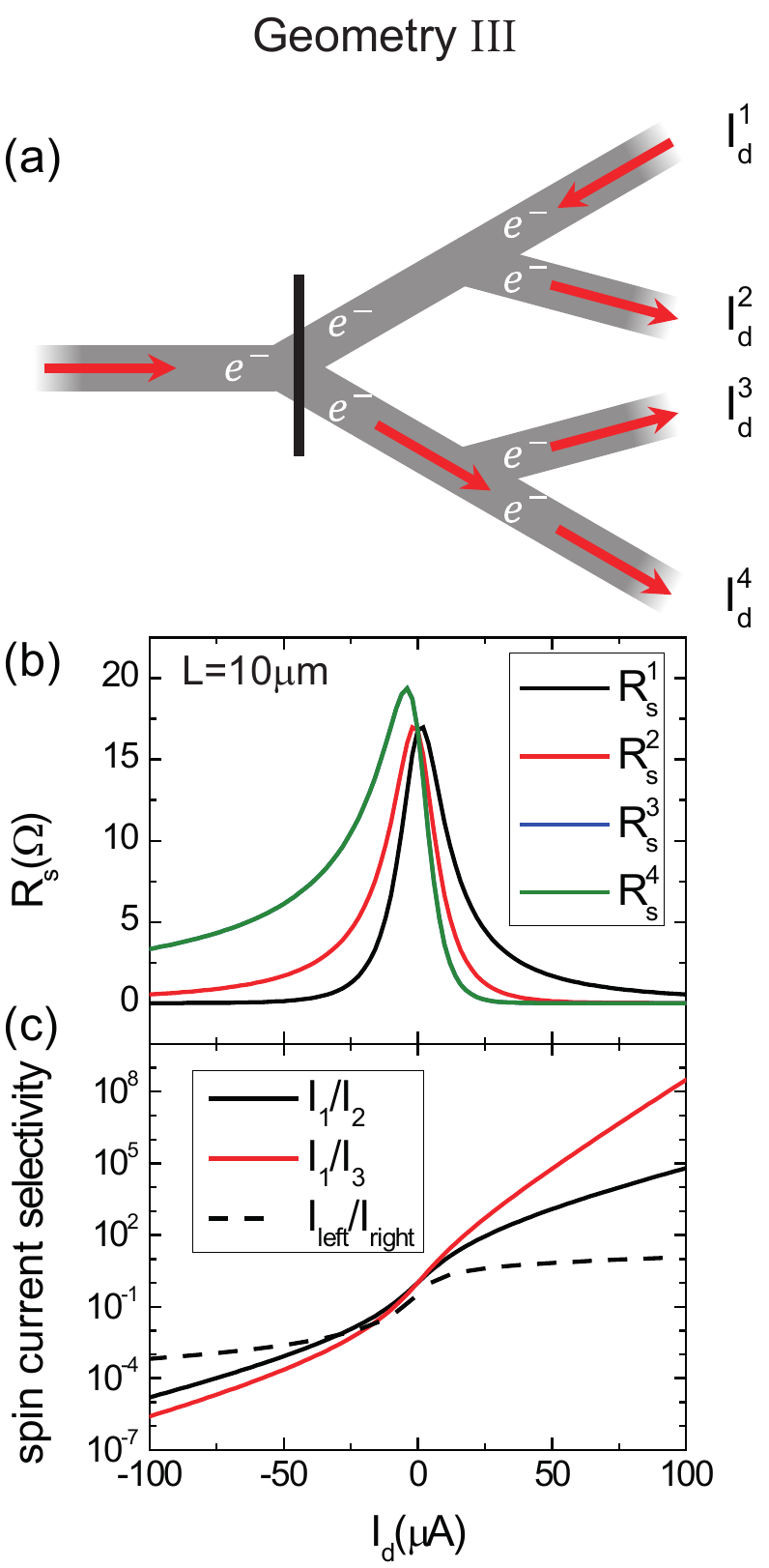}
	\caption{(a) Sketch of device geometry III with the spin currents propagating to output 1 for positive $\mathrm{I_d}$ in the direction of the red arrows. Note that the drift velocity for holes is in the same direction as the drift current. The carriers are electrons in all the arms, the the top and bottom bifurcations are at a distance of $\mathrm{x_1=5}$~\textmu m from the spin injector, the detectors at $\mathrm{L=10}$~\textmu m, and the arms are assumed to be infinitely long. (b) $\mathrm{R_{s}}$ vs $\mathrm{I_d}$ in arms 1 to 4 at $\mathrm{L}$. $\mathrm{R_{s}^3}$ and $\mathrm{R_{s}^4}$ are identical through all the range. (c) Spin current selectivity between arms 1 and 2 and 1 and 3 calculated at $\mathrm{x=L}$ and $\mathrm{I_{left}/I_{right}}$ is calculated at $\mathrm{x=0}$.}
	\label{fig:figure-3}
\end{figure}
\subsection{Geometry III}

Having understood the effect of drift in single Y-shaped graphene devices, we are interested in the operation of graphene channels with several Y-shaped devices connected in series for more complex device functionality.\\
\begin{figure}[tb]
	\centering
		\includegraphics[width=0.4\textwidth]{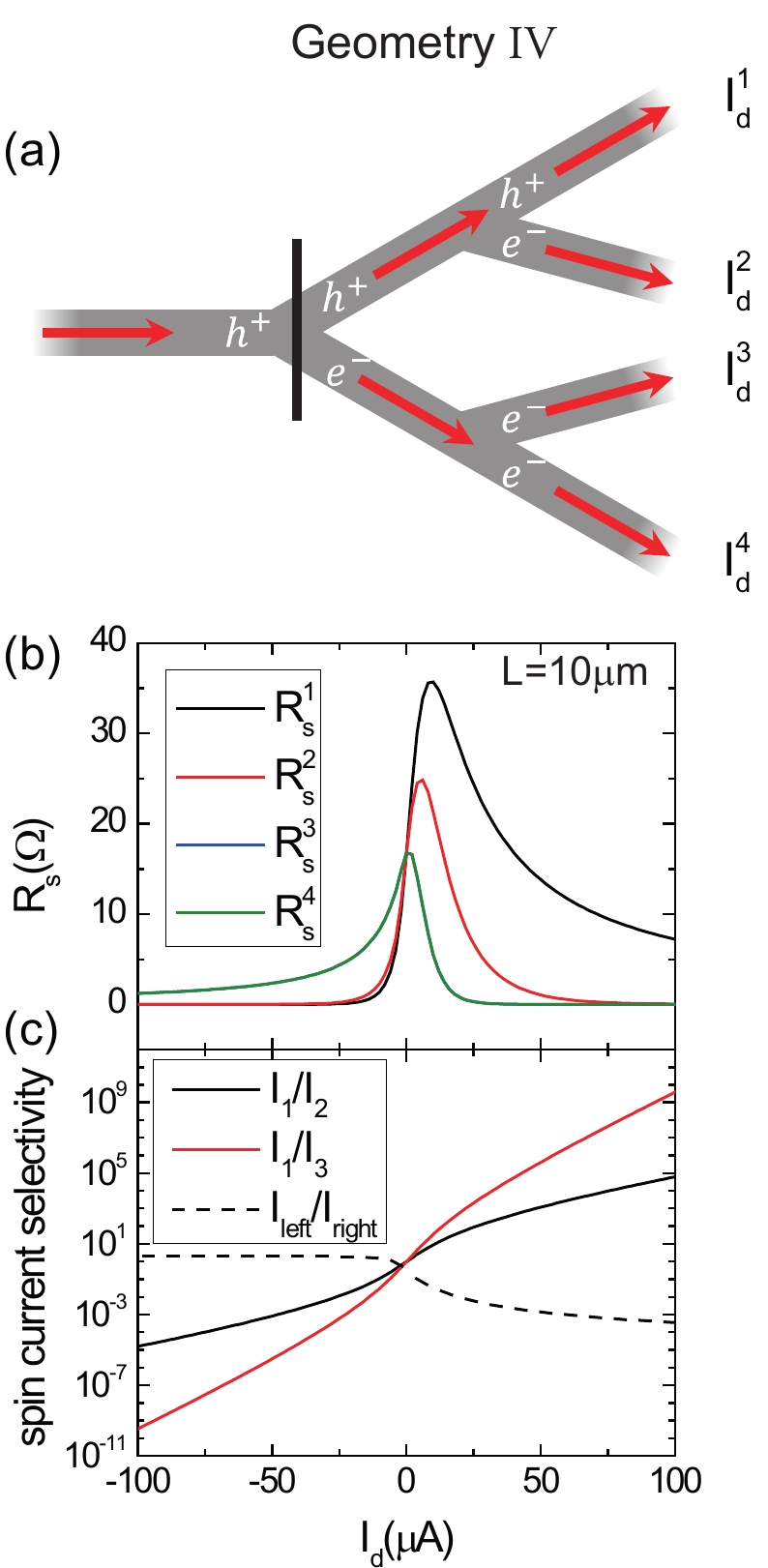}
	\caption{(a) Sketch of device geometry IV, with the spin currents propagating to output 1 for positive $\mathrm{I_d}$ in the direction of the red arrows. The carriers are electrons everywhere apart from the top arm, the top and bottom bifurcations are at a distance of $\mathrm{x_1=5}$~\textmu m from the spin injector, the detectors at $\mathrm{L=10}$~\textmu m, and the arms are assumed to be infinitely long. (b) $\mathrm{R_{s}}$ vs $\mathrm{I_d}$ in arms 1 to 4 at $\mathrm{L}$. $\mathrm{R_{s}^3}$ and $\mathrm{R_{s}^4}$ are identical through all the range. (c) Spin current selectivity between arms 1 and 2 and 1 and 3 calculated at $\mathrm{x=L}$ and $\mathrm{I_{left}/I_{right}}$ is calculated at $\mathrm{x=0}$.}
	\label{fig:figure-4}
\end{figure}
For this purpose, we design a device geometry which is made out of two Y-shaped graphene channels that are connected to the outputs of the first Y-shaped graphene channel. The distance between the bifurcation point where the spin injector is placed and the bifurcation point of the other two is 5~\textmu m (Fig.~\ref{fig:figure-3}a). Using the model described above, we calculate the spin accumulations and spin currents required to understand the performance of these devices. The results are shown in Fig.~\ref{fig:figure-3} for a homogeneous device with the parameters shown in table \ref{table1}. In this case, the nonlocal resistance also decreases for high drift currents. The spin current selectivity between arms 1 and 2 ($\mathrm{I_1/I_2}$) is lower than between arms 1 and 3 ($\mathrm{I_1/I_3}$). This can be explained taking into account that arms 1 and 2 share a 5~\textmu m long channel where there is no drift current, whereas, the drift in the bottom arm is $2\times\mathrm{I_d}$ and opposes spin propagation. When looking at the spin current propagating the left arm we see that it increases for positive drift currents. This is caused by the fact that the drift current in this arm is $2\times\mathrm{I_d}$ and it promotes spin propagation away from the injector at positive drift currents. This not efficient since at $\mathrm{I_d}=100$~\textmu m the spin current propagating towards the left is 11 times higher than the one propagating towards the right and, hence, less than 10\% of the injected spins contribute to the operation.\\
A solution to this issue is to compensate for the current in the left arm by applying a higher drift current to arm 1 $\mathrm{I_d^1}>\mathrm{I_d^2}+\mathrm{I_d^3}+\mathrm{I_d^4}$ while keeping $\mathrm{I_d^2}-\mathrm{I_d^4}$ high enough to prevent propagation in arms 2-4. This is not very efficient because it would lead to an increased power consumption.

\subsection{Geometry IV}

 Alternatively, the spin current propagating in the left arm can be reduced by changing the carrier density in some selected arms. This allows us to apply all four drift currents in the same direction.

The results for such device geometry are reported in Fig.~\ref{fig:figure-4}. In this case, to get the most efficient operation, the charge carriers in the left and top arms are holes whereas the other parts of the sample are electron-doped (see Fig.~\ref{fig:figure-4}a). We see that, in this case, there is a substantial increase in the maximum $\mathrm{R_{s}}$ which can be measured in arm 1 for positive $\mathrm{I_d}$. In this situation, most of the spins propagate towards the top arm with spin current selectivity values up to $5.4\times 10^4$ between arms 1 and 2 and $3.9\times 10^9$ between arms 1 and 3. In this case, the spin currents propagating to the left are $3.5\times 10^{-3}$ times smaller than the ones propagating to the right. This efficiency is provided by the drift current which is $4\times\mathrm{I_d}$ and opposes propagation to the left arm. We also note that, in this case, there is drift in all the arms and, as a consequence, propagation does not rely on slow diffusion. This can be beneficial for the device performance since it enables faster operations.\\ 
We are also interested in the device performance at lower drift currents. In particular, for $\mathrm{I_d}= 26$~\textmu A, $\mathrm{I_{1}/I_{2}}=99$, $\mathrm{I_{1}/I_{3}}=2.6\times 10^3$ and $\mathrm{I_{left}/I_{right}}=5.4\times 10^{-3}$ which we believe should suffice for practical purposes.

 \begin{table*}[tb]
    \caption{Performance for device geometry I-IV. `on/off' is defined as $\mathrm{I_{top}/I_{bot}}$ for geometry I and II. In the case of geometry III and IV `on/off'=$\mathrm{I_1/I_2}(\mathrm{I_1/I_3})$.}
        \begin{ruledtabular}
        \begin{tabular}{c c c c c c c c c}
            \renewcommand{\arraystretch}{2}
&\multicolumn{2}{c}{Geometry I}&\multicolumn{2}{c}{Geometry II}&\multicolumn{2}{c}{Geometry III}&\multicolumn{2}{c}{Geometry IV}\\
\hline
$\mathrm{I}_\mathrm{d}$(\textmu A)&20&100&20&100&26&100&26&100\\
`on/off'& 127&1.1$\times 10^7$ & 127&1.1$\times 10^7$ &99(614) &6.2$\times 10^4$(3.1$\times 10^8$)& 99(2.6$\times 10^3$)&6.2$\times 10^4$(3.9$\times 10^9$)\\
$\mathrm{I_{left}/I_{right}}$&0.23 &0.1 &3.1$\times 10^{-2}$ &1.4$\times 10^{-3}$ &4.1 &12&5.4$\times 10^{-3}$&3.5$\times 10^{-4}$
             
    \end{tabular}
    \end{ruledtabular}
    \label{table3}
\end{table*}
\section{Discussion}
\subsection{Comparison with other approaches} 

As mentioned above, there are alternative methods to control the spin currents in graphene. In Ref.~\cite{MuxFert} the authors show that, due to the change of the spin transport parameters with the carrier density, one can realize demultiplexer functionality in a Y shaped device with local gates addressing each arm individually. This approach has two drawbacks with respect to our approach. Firstly, spin transport in this case is of diffusive nature and, therefore, operation is slower as compared to the drift-based devices. Secondly, tuning the local gates gives substantially smaller contrast between `on' and `off' states as compared to the drift case. This is caused by the relatively low tunability of spin transport parameters in graphene.

Another alternative is to control spin absorption from graphene to an adjacent transition metal dechalcogenide material \cite{yan2016}. `On/off' ratios up to 60 have been demonstrated experimentally and can be further enlarged in high quality devices. However, implementation of this technique for demultiplexer operations, that relies on slow diffusive transport, would result in an overall major suppression of the spin signal that reaches the targeted arm. This is caused by the proximity induced spin-orbit coupling in the graphene channel \cite{Ghiasi2017,benitez2018} that reduces the spin lifetime. Consequently, this approach leads to reduced spin relaxation lengths that require an increase of the input power to achieve signals of the same magnitude.

\subsection{Multiplexer operation and device optimization}

For geometries I and II the use of drift also allows for multiplexer operation. This can be achieved placing two spin injectors in the right side of the sketches in Fig.~\ref{fig:figure-1} or \ref{fig:figure-2}a at a distance significantly longer than the upstream spin relaxation length $\lambda_+$. The modulation of the spin relaxation length induced by drift enables one to select the input that will determine the spin current at the bifurcation point.

 
The main limitation of the drift approach for spin-based demultiplexer and multiplexer operations is the power consumption, which is caused by the drift currents applied in the channel. We suggest two different ways to overcome this issue. The first approach is to increase the distance at which the detectors are placed. This results in higher contrast between `on' and `off' states for lower drift currents. However, this approach also results in lower spin signals and currents because relaxation occurring in the channel reduces the spin accumulation in an exponential way. Since the drift velocity is inversely proportional to the carrier density, reducing this parameter leads to higher power efficiencies. This can be achieved by using semiconducting channels \cite{Kikkawa1999, SpinTransportSiNature}. In particular, black phosphorous is a 2-dimensional semiconductor in which spin lifetimes in the nanosecond range have been reported up to room temperature \cite{BlackPhosphorousSpinInj}. Such lifetimes, together with its high electronic mobilities \cite{AvsarBPContacts} and drift velocity saturations of up to 1.2$\times 10^5$~m/s at room temperature \cite{chen2018large}, make black phosphorous a promising material for spin drift-based devices. 
\section{Conclusions}
In conclusion, we have shown that, by applying a drift current to a Y-shaped graphene-based device, spin currents can be directed in a highly efficient way (see Table~\ref{table3}), allowing for a practical realization of a simple spin-based demultiplexer. In particular, for drift current densities of 20~\textmu A/\textmu m, spin current selectivities up to $10^2$ can be achieved in a 10~\textmu m long device. We have also performed calculations for a device geometry that consists of two Y-shaped graphene channels connected to the outputs of another Y-shaped device, mimicking a demultiplexer with one input and four outputs, and obtained that a similar performance can be achieved by applying drift current densities of 25~\textmu A/\textmu m. We also explain how the effect of drift can be used to achieve multiplexer operation and argue that the introduction of drift is favorable to increase the device operation speed.\\
We believe this is a relevant step forward towards a new generation of spintronic devices with different functionalities.

\section{Acknowledgments}
We acknowledge T. J. Schouten, H. M. de Roosz, H. Adema and J. G. Holstein for technical assistance and J. C. Leutenantsmeyer, T. S. Ghiasi and J. Peiro for discussions. The research leading to these results has received funding from the People Programme (Marie Curie Actions) of the European Union's Seventh Framework Programme FP7/2007-2013/ under REA grant agreement n$^{\circ}$607904-13 Spinograph, the European Union’s Horizon 2020 research and innovation programme under grant agreements No 696656 and 785219 (Graphene Flagship Core 1 and Core 2), NanoNed, the Zernike Institute for Advanced Materials, and the Spinoza Prize awarded to B. J. van Wees by the ‘Netherlands Organization for Scientific Research’ (NWO).

\bibliography{bibliography}
\end{document}